\documentclass[prl,twocolumn,showpacs,floatfix]{revtex4}

\usepackage{times,txfonts,graphicx}
 
\begin{document}
 
\title{Thermodynamics of an interacting trapped Bose-Einstein gas \\ in 
the classical field approximation}

\author{Krzysztof G{\'o}ral$^{1}$, Mariusz Gajda$^{2}$, and Kazimierz 
Rz{\c a}\.zewski$^{1}$}
 
\affiliation{$^1$ Center for Theoretical Physics, $^2$ Institute of 
Physics, \\ Polish Academy of Sciences, Aleja Lotnik\'ow 32/46, 02-668 
Warsaw, Poland}
 

\begin{abstract}
We present a convenient technique describing the condensate in dynamical 
equilibrium  with the thermal cloud, at temperatures close to the critical 
one. 
We show that the whole isolated system may be viewed as a single 
classical field undergoing nonlinear dynamics leading to a steady state.
In our procedure it is the observation process and the finite detection 
time that allow for splitting the system into the condensate and the 
thermal cloud.
\end{abstract}
 

\maketitle


Successful experimental realization of Bose-Einstein condensation in a 
dilute gas of alkali atoms \cite{BEC} offers a new and fascinating tool to 
probe the border between quantum and classical worlds. The theory of the 
dynamical behavior of such a many body system is very difficult and cannot 
be solved exactly. It is particularly hard to extract quantitative 
predictions in the vicinity of the critical temperature. At the other 
extreme, at zero temperature, the weakly interacting Bose gas is well 
described by the mean-field approach. All particles occupy the same 
quantum state whose wave function is the lowest energy solution of the 
Gross-Pitaevskii (GP) equation \cite{GP}. 
At low temperatures the Bogoliubov 
approximation comes handy \cite{Bogol} with its quasi-particles that have 
just been observed in a direct experiment \cite{exp-Bogol}. The theory 
gets much more complex at higher temperatures. There is a considerable 
effort to develop a working theoretical and numerical tool valid there. 
One group of papers \cite{Gora,2quant,correl,stoch,two-gas} from the 
very beginning  describes the system as consisting of two 
(interacting) fractions: the condensate and the thermal cloud.
This ambitious program is very demanding numerically. The other group 
interprets the high energy solutions of the time-dependent GP equation as 
describing the full condensate plus thermal cloud system 
\cite{semiclassical}. As a rule, these authors were able to identify the 
condensate as a definite part of the system only in the somewhat academic 
case of the gas in a rectangular box with periodic boundary conditions. In 
this case 
the condensate is just the zero momentum component of the wave function. 
The aim of this Rapid Communication is to show that much more may 
be achieved with the readily available high-energy solutions of the GP 
equation in harmonic traps. It is also possible to split this solution 
into a sum of the condensed and uncondensed parts, define the condensate 
wave function and approximately estimate the temperature of the resulting 
system.


The Hamiltonian H of the system takes the form:
\begin{eqnarray}
H &=&\int {\rm d^3} r \, \hat \Psi^{\dagger}({\bf r},t)
(\frac{p^2}{2m}+\frac{1}{2}m\omega^{2}r^{2}) \hat \Psi({\bf r},t) 
\nonumber \\
&&+  \frac{2 \pi \hbar^{2} a_{s}}{m} \int {\rm d^3} r \;
\hat \Psi^{\dagger}({\bf r},t) \hat \Psi^{\dagger}({\bf r},t)
\hat \Psi({\bf r},t) \hat \Psi({\bf r},t) \, ,
\end{eqnarray}
where $\hat {\Psi}({\bf r},t)$ is a field operator 
that destroys a particle 
at position {\bf r} and obeys standard bosonic commutation relations 
$[\hat {\Psi}({\bf r},t),\hat {\Psi}^{\dagger}({\bf r'},t)]=\delta({\bf
r}-{\bf r'})$. 
The first term describes particles with mass $m$ trapped in a potential
of a spherically symmetric harmonic oscillator of frequency $\omega$ while
the second term describes two-body interactions.
Here we have assumed that particles interact via a contact 
potential $V({\bf r}-{\bf r'})=4 \pi \hbar^{2}
a_{s}\delta({\bf r}-{\bf r'})/m$, where $a_s$ is the s-wave 
scattering length. 
The Heisenberg equation originating from this Hamiltonian acquires 
the following form
\begin{eqnarray} 
\label{field_dyn}
i\hbar\frac{\partial \hat {\Psi}({\bf r},t)}{\partial 
t}&=&(-\frac{\hbar^{2}\nabla^{2}}{2m}+\frac{1}{2}m\omega^{2}r^{2})\hat 
{\Psi}({\bf r},t)\nonumber\\
&&+\frac{4 \pi \hbar^{2} a_{s}}{m} \hat 
\Psi^{\dagger}({\bf 
r},t) \hat \Psi ({\bf r},t) \hat \Psi({\bf r},t) \; .
\end{eqnarray}
A full operator solution of the non-linear Eq.(\ref{field_dyn}) is not 
known. In addition, the identification of a 
condensate phase is also a subtle issue. The only exception is a system of 
particles in a periodic box. Here the symmetry of the problem helps: 
natural eigenmodes of the system are plane waves with a quantized 
momentum ${\bf k}=2\pi(j_1,j_2,j_3)/L$  ($j_i$ being integer, $L$ - 
box size). Therefore, 
the field operator can be expanded in these modes:
\begin{equation} \label{plane}
\hat \Psi({\bf r},t)=\frac{1}{L^{3/2}}\sum_{{\bf k}}\exp(-i{\bf k}
\cdot{\bf r}) \hat a_{{\bf k}}(t) \; .
\end{equation}
A Bose-Einstein condensate can be uniquely associated with the zero 
momentum mode, ${\bf k}=0$. The annihilation and creation operators $\hat 
a_{{\bf k}}, \hat a_{{\bf k}}^{\dagger}$ satisfy non-linear  equations 
following from Eq.(\ref{field_dyn}).

Different approximate approaches mentioned above
\cite{Gora,2quant,correl,stoch,two-gas} have been introduced in order to 
solve
the problem. Here we want to utilize a method which has been  extensively 
and successfully explored in quantum
optics. A quantum field operator describing a coherent electromagnetic
field (such as laser light) can be  replaced by a complex valued classical
field. Such a substitution is justified for these modes that are highly
occupied. Only then are quantum fluctuations negligible and the
non-vanishing commutator may be ignored. This kind of approximation has
been used recently in \cite{semiclassical} for the matter field: $\hat
a_{\bf k} \rightarrow a_{\bf k}$. Note that $\sum_{\bf k}|a_{\bf k}|^2 = 
N$ and the condensate occupation is equal to $n_0 = |a_0|^2$. On the other
hand, it follows from Eqs.(\ref{field_dyn})-(\ref{plane}) that the
semi-classical approximation is equivalent to a substitution: ${\hat
\Psi({\bf r},t)} \longrightarrow \sqrt{N} \Psi({\bf r},t)  =
\frac{1}{L^{3/2}} \sum_{\bf k}  
a_{\bf k} \exp(-i{\bf k} \cdot{\bf r})$, where $\Psi({\bf r},t)$  fulfills
the  standard GP equation. According to conventional wisdom, this equation 
describes a
pure condensate at zero temperature, hence the whole classical field
$\Psi({\bf r},t)$ represents the condensate populated by $n_0=N$ particles
\cite{Marshall,Adams}.

 

\begin{figure}[h]
\includegraphics[width=\columnwidth,clip]{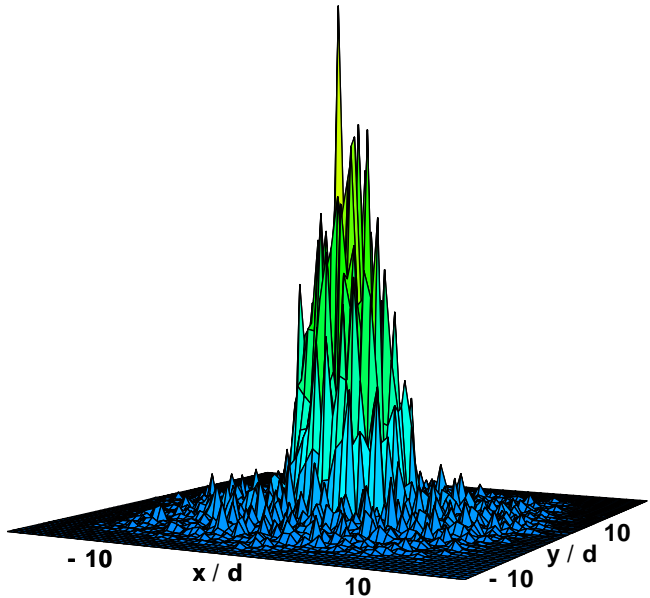}
\caption{\label{fig1}Cross-section of the instantaneous snapshot of the
density distribution in the $z=0$ plane of the Bose gas with the total
energy $E=70 \, \hbar\omega$.
In this and all the following figures lengths are given in units of
$d=\sqrt{\hbar/m\omega}$.
}
\end{figure}

It seems that the two contradicting interpretations of the 
high-energy solution coexist in the literature. In the following we will 
clarify this seeming incongruity.
We focus our 
attention on experimentally relevant system 
of 100 000 $^{87}$Rb atoms trapped in the spherically symmetric harmonic 
potential of frequency $\omega=2 \pi 100$Hz. Our procedure is as follows. 
First, 
we define the wave function of the system on a spatial three-dimensional 
grid. Initial values of the wave function at each point are chosen in 
accordance with the constraints of a fixed energy and particle number. 
Next, we propagate 
this state in time, solving the time dependent GP equation with the help 
of the Fast Fourier Transform. It occurs that for the given total energy 
and particle number  the same steady state is reached after several 
milliseconds. We have checked that the steady state attained does not 
depend on the choice of the initial wave function (in particular, on the 
fact whether its phase is random or not) as long as the number of 
particles and the total energy are kept constant.

\begin{figure}[h]
\includegraphics[width=2\columnwidth,clip]{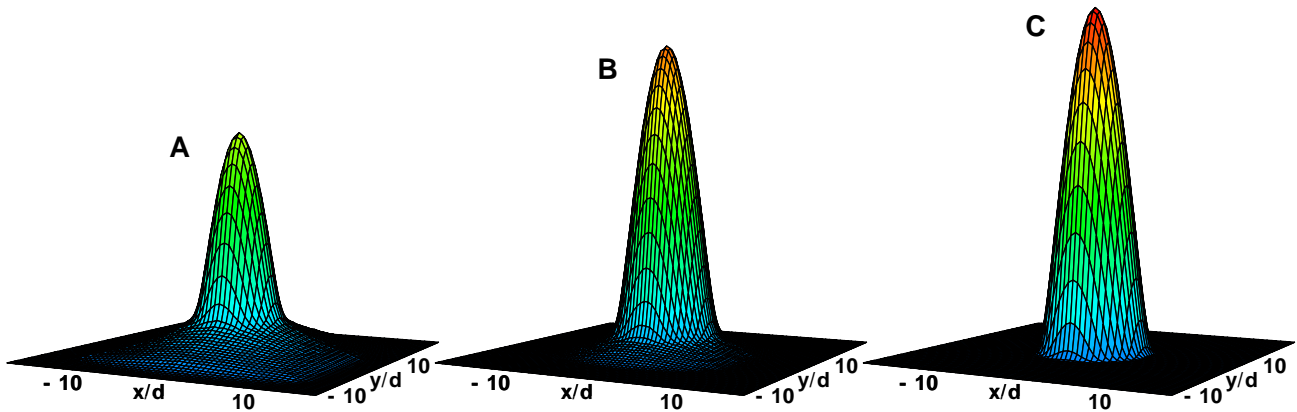}
\caption{\label{fig2}
Time-averaged stationary column density distribution for a Bose gas
plotted for different  values of total energy $E$.
{\bf A} $E=70.00 \, \hbar\omega$ (30\% atoms in condensate). {\bf B}
$E=23.67 \, \hbar\omega$  (72\% of atoms in condensate).  {\bf C} $E=13.04
\, \hbar\omega$  (pure condensate).
}
\end{figure}

The high-energy solutions of the GP equation have striking features 
\cite{Marshall}. A snapshot of the density distribution is shown in 
Fig.\ref{fig1}. The density is 
extremely irregular and exhibits a number of sharp spikes changing its 
shapes and 
positions on very short time scales. Typical methods used to monitor 
trapped atomic condensates involve optical techniques. 
A condensate is illuminated by 
laser light which, after passing through the atomic system (and some 
optical elements), is monitored by a CCD camera. The exposure time $\Delta 
t$
varies from a few microseconds to hundreds of milliseconds and within this 
time the condensate density undergoes rapid changes due to the fast 
dynamics but also because of quantum fluctuations. The nature of the 
observation process introduces a kind of smearing. In fact, in such a long 
exposure a time-averaged rather then an instantaneous single-particle 
density is monitored. The existence of short time 
scales in non-linear many-body dynamics has to be taken into account. 
The object of physical significance is therefore a time-averaged 
single-particle density matrix:   
\begin{equation}
\overline{\rho({\bf r}_1, {\bf r}_2,t)} = \frac{1}{\Delta t} 
\int_{t-\Delta t/2}^{t+\Delta t/2} \Psi^*({\bf r_1},t')\Psi({\bf r_2},t') 
dt' \; .
\end{equation}
This coarse graining procedure destroys the purity of a state of the 
system. 
After the time averaging all irregularities of the density are being 
smoothed out especially in the inner 
part of the density profile (see Fig.\ref{fig2}). The resemblance to
the well publicized 
photographs of the experimental condensates at intermediate temperatures 
is striking \cite{3peaks}. Varying the energy of the gas we can scan the 
whole range
from the pure condensate all the way to 
a nearly critical condition of no condensate. The sample profiles of the 
time-averaged column density (integrated along the z-axis)
are shown in Fig.\ref{fig2}. In parts A and B of 
Fig.\ref{fig2} note the 
bimodal structure of the distribution: the central peak corresponding to 
the Bose-Einstein condensate and the broad background identified with 
the thermal cloud.


We can provide a quantitative analysis of the 
resulting averaged state of the Bose gas. To this end we recall a 
classical definition of the condensate by means of the spectral 
decomposition of the single-particle density matrix \cite{definition}. 
Diagonalization of a single-particle time-averaged density matrix leads 
to natural eigenmodes $\psi_{i}({\bf r},t)$  and eigenvalues ${n_i/N}$:  
\begin{equation}
\overline{\rho({\bf r}_1, {\bf r}_2, t)} = \sum_i {n_i \over N}
\psi^{\star}_{i}({\bf r}_1,t) \psi_{i}({\bf r}_2,t).
\end{equation}
The system may be viewed as a mixture of many coupled coherent 
modes $\psi_i$  whose  occupation is $n_i$. The self-consistency of  the 
model requires that population of each single  mode $\psi_i$ is large, 
$n_i >1$. Otherwise the substitution of the field operator $\hat 
{\Psi}({\bf r},t)$ by a wave function ${\Psi}({\bf r},t)$  would not 
be justified. The existence of the dominant eigenvalue $n_0/N$  of the 
order of unity signifies the presence of the condensate with the 
corresponding eigenvector $\psi_0$  being its wave function.
As the modes and their occupation are known only after the 
time averaging, the verification of the semi-classical criterion can only 
be done {\it a posteriori}. We try a number of grids and choose the one 
that yields the highest occupation of all modes, justifying the 
semi-classical approximation.



The time averaging followed by a diagonalization of  the 
density matrix  gives the spatial modes of the system. These modes are 
typically not known even in a stationary case. Still, the only exception 
is the box with periodic boundary conditions. Let us illustrate our 
method using again this simple case. 

\onecolumngrid

\twocolumngrid

\begin{figure}[h]
\includegraphics[width=\columnwidth,clip]{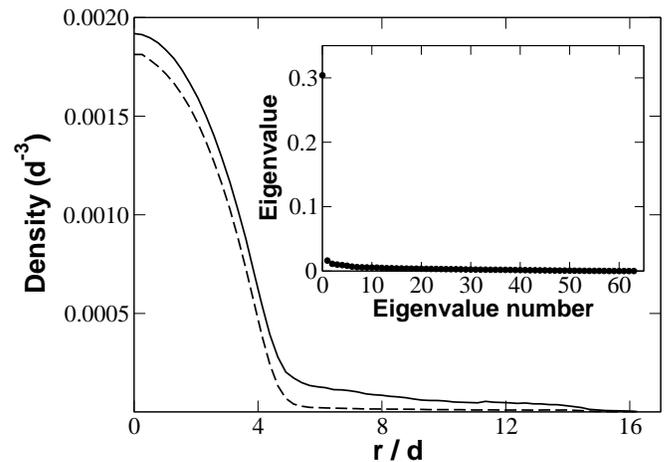}
\caption{\label{fig3}
Time-averaged radial density profile of a stationary state of a Bose
gas. The total energy is $E=70.00 \, \hbar\omega$. Density of the whole
system -- solid line; profile of the Bose-Einstein condensate (defined as
a dominant eigenmode of $\rho_{00}(r_1,r_2)$) -- dashed line.
All eigenvalues of $\rho_{00}(r_1,r_2)$ are shown in the inset.
}
\end{figure}

\noindent Substitution of all annihilation 
operators by complex amplitudes ${\hat a}_{\bf k} \rightarrow a_{\bf k}$ 
gives the following expression for the single-particle density matrix:
\begin{equation}
\overline{\rho({\bf r}_1, {\bf r}_2, t)}   
\approx \frac{1}{V}\sum_{\bf k}
exp[i{\bf k}\cdot({\bf r}_1-{\bf r}_2)] 
|a_{\bf k}(t)|^2
\end{equation}
and the time averaging leaves only diagonal elements ${\bf k}={\bf k'}$ as 
the off-diagonal ones oscillate rapidly and get rapidly dephased,
$\overline{a(t)^*_{\bf k}a(t)_{\bf k'}} \approx \delta_{{\bf k},{\bf k'}}
|a_{\bf k}(t)|^2$. 
Obviously, the $|a_{\bf k}|^2$ are populations of different modes. Only if 
one introduces a kind of coarse graining leading to a suppression of the 
off-diagonal elements of the single-particle density matrix can one 
identify the ${\bf k}=0$ mode as a Bose-Einstein condensate with an 
occupation given by $|a_0|^2$ . Without this additional assumption the 
whole complex field $\Psi({\bf r},t)$  describes one coherent, dynamically 
evolving Bose-Einstein condensate without any thermal cloud. This kind of 
interpretation is used in \cite{Marshall,Adams}. On the contrary, 
identification of the ${\bf k}=0$ momentum component of $\Psi({\bf r},t)$ 
with a Bose condensate has been directly assumed in \cite{semiclassical}.
Our analysis solves this apparent contradiction. By examining a detection 
process and the relevant time scale we can uniquely determine the 
condensate fraction, its wave function as well as 
the structure of excited 
modes of the interacting system. Moreover, the method is no longer 
limited to an academic problem of the uniform system. A steady-state 
single-particle density matrix for the spherically symmetric trapping 
potential must have eigenvectors that are simultaneously diagonalizing 
the angular momentum operators.
Therefore, 
the eigenvectors must be proportional to the spherical harmonics $Y_{lm}$. 
What remains is the one dimensional diagonalization in the radial variable 
of the following projection of the density matrix:
\begin{equation}
\rho_{lm}(r,r')=\int \overline{\rho(r,\Omega; r', \Omega')} \,
Y_{lm}^{\star}(\Omega) \, Y_{lm}(\Omega') \, d\Omega \, d\Omega' \, ,
\end{equation}
where the integration is performed over solid angles $\Omega$ and 
$\Omega'$ associated with the corresponding particle coordinates. We 
expect the condensate to be present in the zero angular momentum component 
of the single-particle density matrix. Indeed, in the inset of 
Fig.\ref{fig3} we show a typical distribution of the eigenvalues of the 
$l=0$ part of the density matrix with one dominant eigenvalue and with the 
corresponding eigenfunction plotted in Fig.\ref{fig3} together with the 
whole time-averaged radial density distribution.

\begin{figure}[hbp]
\includegraphics[width=\columnwidth,clip]{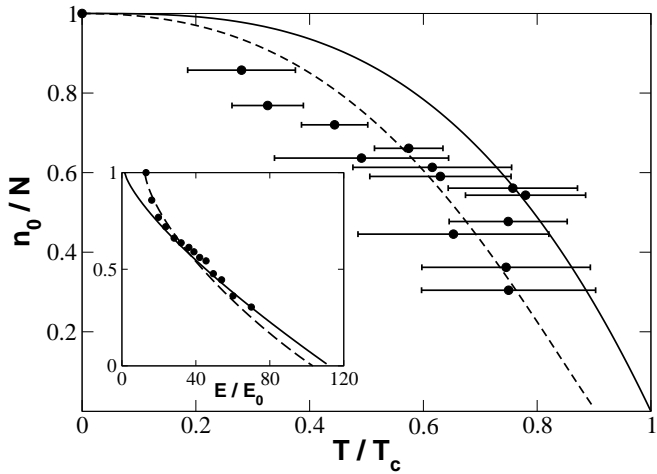}
\caption{\label{fig4}Condensate fraction versus temperature or (the inset)
the total energy of the system. The solid line represents the ideal gas in
the thermodynamic limit, the dashed line depicts the results of \cite{RMP}
obtained within a finite-temperature Hartree-Fock scheme, while dots show
our data. Error bars indicate a range of temperatures for which we obtain
acceptable fits of the ideal Bose gas distribution to the outer wings of
the thermal cloud. $T_c$ is the ideal gas critical temperature in the 
thermodynamic limit and $E_0=\hbar\omega$.}
\end{figure}

The last point raised here is the question of temperature. In principle 
we could do something similar to a typical experiment: switch off the 
trapping potential, let the gas expand and analyze the properties of the 
thermal part. Numerical limitations and the constraining condition of 
the high occupation of each eigenmode make it hard. Alternatively, we 
can follow the procedure of \cite{Varenna} and fit the profile of  the 
thermal fraction of 
the ideal Bose gas to the outer part of our averaged density. This yields 
a reasonable, order of magnitude result. The temperature dependence of the 
condensed fraction together with its large uncertainty is shown in 
Fig.\ref{fig4}.
In the inset we plotted the energy dependence of the number of atoms 
in the condensate. We compare our results with the ideal gas and with a 
two-gas estimate (finite-temperature Hartree-Fock scheme \cite{RMP}). We 
see a reasonably good agreement between all three calculations. 

To summarize: we have shown a simple way of numerically simulating the 
stationary dynamics of a weakly interacting trapped Bose gas at finite 
temperature, which uses a semi-classical representation of the matter 
field and, in accordance with the experimental procedures, stresses the 
role of a finite exposure time when photographing the condensate. In the 
self-consistent determination of the condensate fraction we rely solely on 
the classic criterion of Onsager and Penrose. This way we avoid an 
arbitrary splitting of the system into the  condensed and thermal 
components from the very beginning.
The method may be used to model the impact of thermal fluctuations on the 
dynamical processes with the condensate, such as solitons or vortices. The 
next step in going beyond the approximations employed in our model is the 
estimation of the influence of quantum corrections to the classical 
fields, in particular the study of the corresponding time scales.

We thank J. Mostowski and G.V. Shlyapnikov for stimulating 
discussions. K.R. is supported by the subsidy of the Foundation for Polish 
Science. K.G. and M.G. acknowledge support by Polish KBN grants 5 P03B 
102 20 and 2 PO3B 078 19, respectively. Numerical calculations have been 
performed using computers at Interdisciplinary Centre for Mathematical 
and Computational Modelling  of Warsaw University.


\end{document}